\documentclass[11pt]{article}
\usepackage{epsfig}

\def\prl{Phys. Rev. Lett.}

\begin{document}
\begin{titlepage}
\hfill {KIAS-P03011}

\hfill {SNUTP/03-003}

\hfill {\today: hep-ph/}

\begin{center}
\ \\
{\Large \bf  A Unified Approach to High Density:\\ Pion
Fluctuations in Skyrmion Matter}
\\
\vspace{.30cm}

Hee-Jung Lee$^{a,b}$, Byung-Yoon Park$^{a,b}$, {Dong-Pil
Min$^{c}$}, Mannque Rho$^{a,d}$ \\ and Vicente Vento$^{a,e}$

\vskip 0.20cm

{(a) \it School of Physics, Korea Institute for Advanced Study,
Seoul 130-012, Korea}

{(b) \it Department of Physics,
Chungnam National University, Daejon 305-764, Korea}\\
({\small E-mail: hjlee@phya.snu.ac.kr, bypark@chaosphys.cnu.ac.kr})

{(c) \it Department of Physics and Center for Theoretical Physics}\\
{\it Seoul National University, Seoul 151-742, Korea}\\
({\small E-mail: dpmin@phya.snu.ac.kr})

{(d) \it Service de Physique Th\'eorique, CE Saclay}\\
{\it 91191 Gif-sur-Yvette, France}\\
({\small E-mail: rho@spht.saclay.cea.fr})

{(e) \it Departament de Fisica Te\`orica and Institut de
F\'{\i}sica
Corpuscular}\\
{\it Universitat de Val\`encia and Consejo Superior
de Investigaciones Cient\'{\i}ficas}\\
{\it E-46100 Burjassot (Val\`encia), Spain} \\ ({\small E-mail:
Vicente.Vento@uv.es})

\end{center}
\vskip 0.3cm

\centerline{\bf Abstract}

As the first in a series of systematic work on dense hadronic
matter, we study the properties of the pion in dense medium using
Skyrme's effective Lagrangian as a unified theory of the hadronic
interactions applicable in the large $N_c$ limit. Dense baryonic
matter is described as the ground state of a skyrmion matter which
appears in two differentiated phases as a function of matter
density: i) at high densities as a stable cubic-centered (CC)
half-skyrmion crystal; ii) at low densities as an unstable
face-centered cubic (FCC) skyrmion crystal. We substitute the
latter by a stable inhomogeneous phase of lumps of dense matter,
which represents a naive Maxwell construction of the phase
transition. This baryonic dense medium serves as a background for
the pions whose effective {\em in-medium} Lagrangian we construct
by allowing time-dependent quantum fluctuations on the classical
dense matter field. We find that the same parameter which
describes the phase transition for baryonic matter, the
expectation value of the $\sigma$ field, also describes the phase
transition for the dynamics of the {\em in-medium} pion. Thus, the
structure of the baryonic ground state $crucially$ determines the
behavior of the pion in the medium. As matter density increases,
$\langle\sigma\rangle$ decreases, a phenomenon which we interpret
to signal, in terms of the parameters of the effective pion
Lagrangian $f_\pi^*$ and $m_\pi^*$, the restoration of chiral
symmetry at high density. Our calculation shows also the important
role played by the higher powers in the density as it increases
and chiral symmetry is being restored. This feature is likely to
be generic at high density although our ground state may not be
the true ground state.

\vskip 0.3cm \leftline{Pacs: 12.39-x, 13.60.Hb, 14.65-q, 14.70Dj}
\leftline{Keywords:  pion, effective mass, dense matter, chiral
symmetry}

\end{titlepage}

\section{Introduction}
 Understanding the properties of hadrons in a dense nuclear
medium is currently an important issue in nuclear physics. The
dynamically generated masses of the light hadrons reflect the
symmetry breaking pattern of QCD. The quark condensate $\langle
\bar{q} q \rangle$ which represents the order parameter drops as
the density of the nuclear medium increases. Consequently all
light-quark hadron masses (except that of the pions) are expected
to follow the behavior of the condensate, a basis of the scaling
proposed in \cite{BR91} and predicted in an effective field theory
with the vector manifestation fixed point~\cite{HY:PR}. The pion
however is very special, because it is the Goldstone boson
associated with the spontaneous symmetry breakdown of the chiral
symmetry. The Hellmann-Feynman theorem~\cite{CFG92,DL91}, a mean
field approximation in chiral perturbation
theory~\cite{BKRT92,YNMK93,YMK94,TW95}, and the Nambu-Jona-Lasinio
model~\cite{BKM87,HK85,RD87} lead to the same linear dependence in
the baryon number density $\rho$ namely,
\begin{equation}
\frac{\langle \bar{q}q \rangle_{\rho}}{\langle \bar{q}q\rangle_{\rho=0}}
= 1 - \frac{\Sigma_{\pi N}}{m_\pi^2 f_\pi^2} \rho + \cdots,
\end{equation}
where $\Sigma_{\pi N}$ is the pion-nucleon sigma term. If we
assume that the Gell-Mann-Oakes-Renner (GMOR) relation holds for
finite density, we have~\footnote{For the moment, we are not
distinguishing the time and space components of the pion decay
constant. Which one we are dealing with will be specified later.}
\begin{equation}
f_\pi^{*2} m_\pi^{*2} \approx - m_q \langle \bar{q} q\rangle_\rho,
\label{HF}\end{equation}
where $f_\pi^*$ is the in-medium pion decay constant and $m_\pi^*$
the effective pion mass in the medium. There has been a lot of
discussions on the physical consequence of eq.~(\ref{HF}) with a
decreasing quark condensate. The decrease in the {\em pion decay
constant} is interpreted as a symptom of partial chiral
restoration in the nuclear medium~\cite{HK85}. On the other hand,
in refs.~\cite{BKRT92,YNMK93,YMK94}, the possibility of S-wave
pion condensation resulting from the decrease of the {\em pion
effective mass} as a consequence of the decrease of $\langle
\bar{q} q\rangle$ in eq.~(\ref{HF}) has been discussed. However,
if one lets the pion decay constant change, then the effective
pion mass can even increase with increasing nuclear density as
discussed in ref.~\cite{TW95} and a completely different scenario
can then arise.

The deeply bound pionic states observed in heavy nuclei (Pb, Sn
etc.) at GSI~\cite{DBPS,Sn} have renewed interest in these
issues~\cite{WBW97,Yama98,FG98,HZC00,PJM02,GNO02,KKW02}. The well
defined 1s and 2p states with narrow width provide a new standard
for the S-wave pion-nucleus optical potential $U_s(r)$, whose real
part may be identified as the effective mass of the
pions~\cite{WBW97,Yama98}
\begin{equation}
m^{*2}_\pi(r) = m^2_\pi
 + 2\omega \mbox{Re}  U_s(r).
\label{mpi_r}\end{equation}
This formula would seem to imply that the pion mass increases a
7\% at the center of the Pb nucleus and a 3\% in symmetric nuclear
matter. But other possibilities exist. For instance, if one
assumes the pion mass to be unmodified for low density which is
not unplausible~\footnote{The pion mass of course remains strictly
unmodified in the chiral limit but it is possible that even in the
presence of small quark masses, the approximate chiral invariance
protects the pion mass so that it is left more or less unaffected
by density, at least at low density. An $SU(2)_c$ (2 color) QCD
lattice calculation indicates that this is what
happens~\cite{nakamura}.}, then the measurements provide
information on the scaling of the pion decay constant instead of
that of the mass. See \cite{Sn}.

Theoretical studies on the effective mass of the pion in a nuclear
medium follow two different schemes. One traditional approach
relies on optical potentials. However, it is difficult to relate
the phenomenological $\pi$-nucleus potential to the microscopic
$\pi$-$N$ interactions in a systematic way. The other more modern
approach proceeds via chiral perturbation theory~\cite{PMR93} with
explicit baryon fields. By integrating out the baryon fields,
using the mean field approximation, one obtains an effective
theory whose parameters become dependent on the nuclear density
$\rho$. These calculations consist of computing single-nucleon
loops leading to a linear dependence on the density of the
parameters. Higher orders in the nucleon loop expansion would
provide us with higher powers of $\rho$, however, due to the
strong couplings associated with the nuclear forces, it is not
clear that the results will converge although chiral perturbation
power counting is used. Perhaps more importantly an effective
field theory Lagrangian properly matched to QCD at an appropriate
scale as discussed by Harada and Yamawaki~\cite{HY:PR} $must$ have
a highly nontrivial ``intrinsic" dependence on density (or
temperature if in heat bath) in the parameters  but this
dependence is generally missing in the chiral Lagrangian so far
used~\cite{TW95,WBW97}.

We study the pion properties in dense baryonic matter in an
approach based on Skyrme's Lagrangian, which we use to describe,
in a unified manner, the dynamics of baryons, pions and the
interactions between them. We proceed by calculating initially the
ground state of infinite skyrmion matter. At high density it is
described by a skyrmion crystal. At lower density the crystal
suffers a phase transition and becomes unstable. The system
chooses then as its ground state an inhomogeneous phase where
skyrmions condense to form lumps of matter of higher density. We
adopt a naive Maxwell construction to describe such state. We
incorporate the pion in this medium by performing quantum
fluctuations in the pion field of the Lagrangian describing this
state. We obtain in this way the effective dynamics of the pion in
dense skyrmion matter. The parameters of this Lagrangian are
dependent on matter density and contain {\em all higher orders in
density} as dictated by Skyrme's Lagrangian.

Before proceeding we must clearly lay down the scope of this work
here. We do not claim that the result we obtain -- which is quite
striking and intriguing -- is in any way reflecting Nature. In
fact it is perhaps likely that the ``ground state" we obtain is
not close to what it should be. More work is needed to obtain from
what we have a true ground state and we have some ideas as to how
to proceed. What we do here is that we will simply assume that we
have certain states given by exact classical solutions of the
field equations of a theory that is considered to be valid at
large $N_c$, namely, the crystal structure of different
symmetries, and study what they imply for the excitations thereon.
This work should be taken as representing the first exploratory
step towards a more realistic treatment -- that is in accordance
with QCD -- of highly dense matter, the objective of our effort.

This paper is organized as follows. In the next section, after a
brief description of the model Lagrangian, we construct the
classical skyrmion crystal solution using Fourier expansions. In
sec.~3, we incorporate the pion fluctuations in the ground state
and obtain the effective pion Lagrangian in the skyrmion medium.
We discuss how the pion decay constant and the effective pion mass
change in matter. Section~4 summarizes the most important results
and contains our conclusions.

\section{The Model Lagrangian and the FCC Skyrmion Crystal}
The description of baryonic matter we are going to investigate is
based on the Skyrme model Lagrangian~\cite{Sk62} for the massive
pion fields, which reads
\begin{equation}
{\cal L}= -\frac{f_\pi^2}{4} \mbox{Tr} (U^\dagger \partial_\mu U
U^\dagger \partial^\mu U) + \frac{1}{32e^2} \mbox{Tr}
 [U^\dagger \partial_\mu U, U^\dagger \partial_\nu U]^2
+ \frac{f_\pi^2 m_\pi^2}{4} \mbox{Tr} (U+U^\dagger -2 ),
\label{Lsk}\end{equation}
where $U\in SU(2)$ is a nonlinear realization of the pion fields.
$f_\pi$, $e$ and $m_\pi$ are the pion decay constant, Skyrme's
parameter and the pion mass, respectively. Note that this
Lagrangian (\ref{Lsk}) is described in terms of $U$, which
contains only pionic degrees of freedom. We will not consider the
isospin symmetry breaking due to the u- and d-quark mass
difference, since it is of no relevance for the present study. Our
aim is to study the properties of the pion in a dense baryonic
medium. Since this is an effective Lagrangian, if matched to
QCD~\cite{HY:PR}, its parameters should depend on the properties
of the medium , i.e. matter density in our case as discussed in
\cite{BR:PR01}. However we follow a rather different philosophy,
namely we shall assume that the above Lagrangian embodies a
complete description of $all$ physical processes of interest,
i.e., free pions, free baryons, many baryon states, dense baryonic
matter and moreover, it also describes the pions interacting with
any of these baryonic systems. We fix therefore the parameters of
the Lagrangians in one of these processes, for example by
describing the free pion and nucleon systems, and calculate how
these values change as the density of the baryonic matter
increases. From the recent understanding~\cite{HY:PR,BR:PR01} of
effective theories that purport to match to QCD, we know that this
is not totally consistent since EFTs can make sense only for given
scales and the parameters of the theory necessarily flow. This
issue which is quite intricate will be dealt with in a forthcoming
publication.

For completeness we should also mention that (\ref{Lsk}) is the
most economical effective Lagrangian that one can take. To be
truly realistic, massive fields -- such as the light-quark vector
mesons, scalar mesons or ``dilatons"~\footnote{Implementing
dilatons as was done in \cite{BR91} will play the role of
incorporating the ``intrinsic density dependence" inherent in
Harada-Yamawaki theory of hidden local symmetry with the vector
manifestation fixed point~\cite{HY:PR}. This point is being
pursued and will be discussed in a forthcoming publication.}
associated with trace anomaly etc -- and/higher derivative terms
need be incorporated but for our purpose it suffices to take the
simplest yet sufficiently realistic form given by (\ref{Lsk}).

A similar procedure has been followed for example in  chiral
perturbation theory with explicit nucleon
fields~\cite{YMK94,TW95}. They start with a Lagrangian containing
all the degrees of freedom and free parameters, they integrate out
the nucleons in and out of an \`a priori assumed Fermi sea and in
the process they get a Lagrangian density describing the pion in
the medium which corresponds to the above Skyrme Lagrangian except
that the quadratic (current algebra) and the mass terms pick up a
density dependence of the form
\begin{equation}
-\frac{f_\pi^2}{4} \left( g^{\mu\nu} +  \frac{D^{\mu\nu} \rho}
{f_\pi^2}\right) \mbox{Tr} (U^\dagger \partial_\mu U U^\dagger
\partial^\nu U)
 + \frac{f_\pi^2 m_\pi^2}{4}
\left( 1 - \frac{\Sigma_{\pi N}}{f_\pi^2 m_\pi^2}  \rho \right)
\mbox{Tr} (U+U^\dagger -2 ), \label{ChPT}\end{equation}
where $\rho$ is the density of the nuclear matter and $D^{\mu\nu}$
and $\sigma$ are physical quantities obtained from the
pion-nucleon interactions. Note that in this scheme, nuclear
matter is assumed ab initio to be a Fermi sea devoid of the
intrinsic dependence mentioned above. We shall discover in the
next section that our free pion Lagrangian will be also modified
in a similar manner but in the crudest approximation. At a more
advanced level, there will be an intricate nontrivial density
dependence missing in the chiral perturbative calculations.

Having stated our scheme we proceed to describe dense baryonic
matter in the Skyrme model. Consider a static field configuration
$U_0(\vec{x})= \sigma +i\vec{\tau}\cdot\vec{\pi}$ with
$\sigma^2+\vec{\pi}^2=1$. It is a nonlinear realization of the
pion fields and a mapping, $S^3$, from  real space $R^3$ to the
internal $SU(2)$ space. If we take for $U_0$ at infinity a
constant SU(2) matrix, the configurations can be classified
according to their winding number
\begin{equation}
B=\int d^3 x \rho_B(\vec{x}),
\end{equation}
with
\begin{equation}
\rho_B(\vec{x})= \frac{1}{24\pi^2} \epsilon_{ijk}
{\rm Tr} (U_0^\dagger \nabla_i U_0
 U_0^\dagger \nabla_j U_0 U_0^\dagger \nabla_k U_0).
\end{equation}
Skyrme~\cite{Sk62} used this winding number to describe the baryon
number. It is therefore through topology that this Lagrangian,
written  in terms of pion fields, describes  baryonic systems.

Various solutions of static field configurations which describe
infinite matter can be obtained from skyrmions by assigning
specific symmetries~\cite{Kl85,BJW85,GM87,JV88,CJJVJ89,KS89}. We
proceed to describe in detail one of these solutions as a first
step to study the behavior of the elementary pion in skyrmion
matter. We adopt the method developed by Kugler and
Shtrikman~\cite{KS89}. They construct as an Ansatz for skyrmion
matter a crystal whose configurations are such that each
neighboring pair is rotated in isospin space, relative to the
other  with respect to the line joining them, by $\pi$. Our
starting crystal is FCC~(face centered cubic) and the local baryon
number takes its maximum at the FCC lattice cites. At high
density, this configuration makes a phase transition to a
CC~(cubic centered) crystal, so called ``{\em
half-skyrmion}"~\cite{GM87} crystal, because only half of the
baryon number carried by the single skyrmion is concentrated at
the original FCC cites, while the other half is concentrated at
the center of the links connecting those points, where the baryon
number density was negligible initially. Thus the phase transition
leads to a CC crystal made up of half-skyrmions.

Let us construct the initial skyrmion matter solution. Consider a
point in space $\vec{x}=(x,y,z)$ at which the field is given by
$(\sigma, \pi_1, \pi_2, \pi_3)$. Then, the FCC configuration is
defined by the following symmetries:
\renewcommand{\theenumi}{\arabic{enumi}}
\renewcommand{\labelenumi}{(S\theenumi)}
\begin{enumerate}
\item under reflection in space $\vec{x} \rightarrow (-x,y,z)$,
the field is also reflected in isospin space according to
$(\sigma, -\pi_1, \pi_2, \pi_3)$, \label{S1}
\item under a rotation around the threefold axis in space
$\vec{x} \rightarrow (y,z,x)$, the field is simultaneously rotated
in isospin space about the corresponding axis in isospin space
according to $(\sigma, \pi_2, \pi_3, \pi_1)$, \label{S2}
\item under a rotation around the fourfold axis in space
$\vec{x} \rightarrow (x,z,-y)$, the field is rotated around the
corresponding fourfold axis in isospin space according to
$(\sigma, \pi_1, \pi_3, -\pi_2)$, \label{S3}
\item under a translation from a corner of a cube to the center
of a face $\vec{x} \rightarrow (x+L,y+L,z)$, the field becomes
rotated by $\pi$ about the axis perpendicular to the face
according to $(\sigma, -\pi_1, -\pi_2, \pi_3)$. \label{S4}
\end{enumerate}
Here, $2L$ is the size of the single FCC unit cell which contains
4 skyrmions. Thus, the baryon number density is $\rho=1/2L^3$. The
normal nuclear matter density $\rho_0 =0.17/$fm$^3$ corresponds to
$L\sim 1.43$ fm.

We initially introduce ``{\em unnormalized}" fields
$(\bar{\sigma}, \bar{\pi}_1, \bar{\pi}_2, \bar{\pi}_3)$, which are
then ``{\em normalized}" to define properly $U_0=\sigma+i
\vec{\tau}\cdot\vec{\pi}$ by
\begin{equation}
\sigma = \frac{\bar{\sigma}}{\sqrt{\bar{\sigma}^2+\bar{\pi}_1^2
+\bar{\pi}_2^2 + \bar{\pi}_3^2 }}, \label{norm}\end{equation}
with similar equations being satisfied by  $\pi_i~(i=1,2,3)$. The
field configurations obeying the above symmetries can be easily
found by expanding the unnormalized fields as Fourier series, i.e.
\begin{equation}
\bar{\sigma} = \sum_{a,b,c} \bar{\beta}_{abc} \cos(a\pi x/L)
\cos(b\pi y/L) \cos(c\pi z /L),
\label{sigma}\end{equation} and
\begin{eqnarray}
\bar{\pi}_1 &=& \sum_{h,k,l} \bar{\alpha}_{hkl} \sin(h\pi x/L)
\cos(k\pi y/L) \cos(l\pi z/L),
\label{pi1} \\
\bar{\pi}_2 &=& \sum_{h,k,l} \bar{\alpha}_{hkl} \cos(l\pi x/L)
\sin(h\pi y/L) \cos(k\pi z/L),
\label{pi2} \\
\bar{\pi}_3 &=& \sum_{h,k,l} \bar{\alpha}_{hkl} \cos(k\pi x/L)
\cos(l\pi y/L) \sin(h\pi z/L).
\label{pi3}
\end{eqnarray}
The symmetry relations (S\ref{S1})-(S\ref{S4}) restrict the modes
appearing in eqs.~(\ref{sigma}-\ref{pi3}) as follows;
\renewcommand{\theenumi}{\arabic{enumi}}
\renewcommand{\labelenumi}{(M\theenumi)}
\begin{enumerate}
\item if $h$ is even, then $k,{}l$ are restricted to odd numbers
and $a,{}b,{}c$ are to even numbers,
\label{M1}
\item if $h$ is odd, then $k,{}l$ are restricted to even numbers
and $a,{}b,{}c$ are to odd numbers.
\label{M2}
\end{enumerate}
Furthermore, the coefficients of the expansion satisfy
$\bar{\alpha}_{hkl}=\bar{\alpha}_{hlk}$ and
$\bar{\beta}_{abc}=\bar{\beta}_{bca}=\bar{\beta}_{cab}
=\bar{\beta}_{acb}=\bar{\beta}_{cba}=\bar{\beta}_{bac}$. Note that
the normalization process (\ref{norm}) does not spoil any
symmetries that the unnormalized fields have, while the expansion
coefficients $\alpha_{hkl}$ and $\beta_{abc}$ lose their meaning
as Fourier coefficients in the normalized fields.

Without loss of generality, we can locate the centers of the
skyrmions  at the corners of the cube and at the centers of the
faces by letting ${\sigma}=-1$ and ${\pi}_i(i=1,2,3)=0$ at those
points. For the skyrmion field to have definite integer baryon
number per cite, we should have ${\sigma}=+1$ and
${\pi}_i(i=1,2,3)=0$ at the points such as $(L,0,0)$. It gives the
constraint
\begin{equation}
\sum_{a,b,c=\mbox{\scriptsize even}} \bar{\beta}_{abc}=0.
\end{equation}

If only the modes (M\ref{M2}) appear in the expansion, the
configuration has an additional symmetry, namely that under the
translation $\vec{x}\rightarrow (x+L,y,z)$, the field rotates
under $O(4)$  by $\pi$ in the $\sigma, \pi_1$ plane. This
configuration corresponds to the {\em half-skyrmion} CC discussed
above. Because of this additional symmetry, the physical
quantities such as the local baryon number density and the local
energy density become completely identical around the points with
$\sigma=-1$ and the points with $\sigma=+1$. Thus, one half of the
baryon number carried by a single skyrmion is concentrated at the
cites where the centers of the single skyrmion is in the FCC
crystal. The other half of the baryon number is now concentrated
on the links connecting those points, where  $\sigma$ takes the
value $+1$, while in this location in the FCC configuration the
baryon number density is low. As a consequence the average value
$\langle \sigma \rangle$ vanishes and we will see that this
phenomenon signals that, in the dense medium, chiral symmetry
restoration for the pion dynamics takes place. It is important to
stress that it is the precise structure of the ground state which
is responsible for the restoration.  Both modes (M1) and (M2) are
included in the calculation. The half-skyrmion crystal arises as
the stable ground state at high density, because the expansion
coefficients associated with the modes (M1) become suppressed.

The Fourier series expansion method works well for the
unnormalized fields and only a few modes are
necessary.\footnote{Of course, through the normalization process
(\ref{norm}), infinitely many modes appear in the normalized
fields. However, the expansions converge rapidly.} The expansion
coefficients are used as variational parameters and determined by
minimizing the energy of the configuration. The coefficients
depend on the box size $L$. In Table~1, we list a few modes below
$E=16(\pi/L)^2$ and we use only the modes with $E \le 10(\pi/L)^2$
in our calculation.

\begin{table}
\caption{A few modes used in the Fourier series expansion
coefficients $\alpha_{hkl}$ and $\beta_{abc}$ in
eqs.~(\ref{sigma}-\ref{pi3}). $E$ is the energy of the modes in
units $(\pi/L)^2$  and $d$ is the degeneracy of the mode.}
\begin{center}
\begin{tabular}{ccccc|ccccc}
\hline
$h$ & $k$ & $l$ & $E$ & $d$ & $a$ & $b$ & $c$ & $E$ & $d$ \\
\hline
 1 & 0 & 0 & 1 & 1 & 0 & 0 & 0 & 0 & 1 \\
 1 & 2 & 0 & 5 & 2 & 1 & 1 & 1 & 3 & 1 \\
 2 & 1 & 1 & 6 & 1 & 2 & 0 & 0 & 4 & 3 \\
 1 & 2 & 2 & 9 & 1 & 2 & 2 & 0 & 8 & 3 \\
 3 & 0 & 0 & 9 & 1 &   &   &   &    &   \\
\hline
 2 & 3 & 1 & 12 & 2 & 3 & 1 & 1 & 11 & 3 \\
 3 & 2 & 0 & 13 & 2 & 2 & 2 & 2 & 12 & 1 \\
\hline
\end{tabular}
\end{center}
\end{table}

In Fig.~1 we show the energy per baryon $E/B$ as a function of the
FCC box size parameter $L$. Each point denotes the minimum energy
of the crystal configuration with a given value of $L$, which we
find by using a proper minimization program such as ``the
down-hill simplex method"~\cite{NR}. The solid circles correspond
to the zero pion mass calculation and reproduce those of Kugler
and Shtrikman~\cite{KS89}. In order for an easier comparison  we
present the results with $L$ given in units of $(ef_\pi)^{-1}$
$(\sim 0.45$fm) and $E/B$ in units of $(6\pi^2 f_\pi)/e$ $(\sim
1160$ MeV), respectively. The latter enable us to compare the
numerical results on $E/B$ easily with its Bogolmoln'y bound for
the skyrmion in the chiral limit, which can be expressed as
$E/B=1$ in this convention.

By looking at the figure one sees that as we squeeze the system
from $L=6$ to around $L =3.8$, the skyrmion system undergoes a
phase transition from the FCC single skyrmion configuration to the
CC half-skyrmion configuration. The system has a minimum energy at
$L\sim 2.4$ with the energy per baryon $E/B\sim 1.038$, which is
very close to the Bogolmoln'y bound. One can note in Fig.~1 that
in the region $L>3.8$, two different phases coexist. This
coexistence leads to an interesting hysteresis phenomenon in the
numerical process for finding the solutions, where one uses the
found solution set for an L as the initial trial functions for the
next L. If one starts from a small $L$, the numerical program
might stay in a {\em quasi-stable} half-skyrmion CC configuration
up to quite large values of $L>6$, then suddenly it changes to the
single skyrmion FCC configuration. This phase transition -- and
hysterisis -- occurs in the unstable region where the system has
negative pressure, so does not have any physical consequence.

The phase to the left of the minimum is referred to in this work
as ``{\em homogeneous}" and there the background fields are
described by a crystal configuration. The phase to the right of
the minimum is ``{\em inhomogeneous}" since the pressure
$P\equiv\partial E/\partial V$ is negative, the skyrmion matter is
unstable, and to make it stable we will assume a realization in
terms of an ``{\em inhomogeneous phase}" where the skyrmions
condense to form dense lumps with large empty spaces.

The open circles are the solutions found with a nonvanishing pion
mass, $m_\pi=140$ MeV.\footnote{Incorporating the pion mass into
the problem introduces a new scale in the analysis and therefore
we are forced to give specific values to the parameters of the
chiral effective Lagrangian, the pion decay constant and the
Skyrme parameter, a feature which we have avoided in the chiral
limit. In order to proceed, we simply take their empirical values,
that is, $f_\pi=93$ MeV and $e=4.75$. Although the numerical
results depend on these values, their qualitative behavior will
not.} Comparing to the skyrmion system for massless pions, the
energy per baryon is somewhat higher. Furthermore, there is no
first order phase transition at low densities.

\begin{figure}
\centerline{\epsfig{file=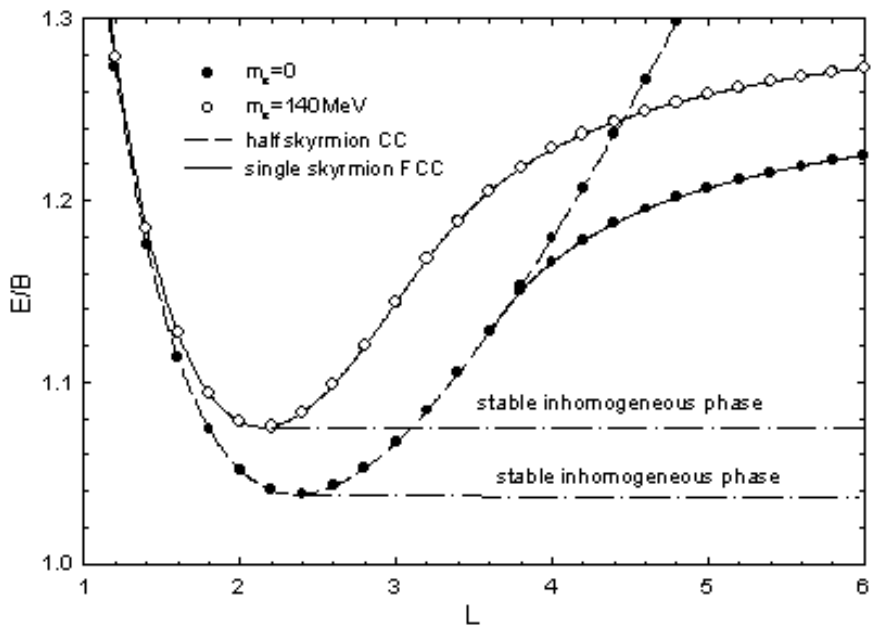,width=12cm,angle=0}}
 \caption{Energy
per single skyrmion as a function of the size parameter $L$. The
solid circles show the results for massless pions and the open
circles are those for massive pions. Note the rapid phase
transition around $L\sim 3.8$ for massless pions.}
\end{figure}

In Fig.~2, we present $\langle \sigma \rangle$, i.e. the average
value of $\sigma$ over space as a function of $L$. In the chiral
limit, $\langle \sigma\rangle$ rapidly drops as the system
shrinks. It reaches  zero at $L\sim 3.8$. where the system goes to
the half-skyrmion phase. This phase transition can be interpreted,
once the pion fluctuations are incorporated, as a signal for
chiral symmetry restoration. In the case of massive pions, the
transition in $\langle \sigma\rangle$ is soft. Its value
monotonically decreases and reaches zero asymptotically, as the
density increases. Furthermore, as we can see in Fig.~3, where the
local baryon number density is shown, for $L=2$ (left) and $L=3.5$
(right) in the $z=0$ plane, the system is (almost) a half-skyrmion
crystal at high density.

\begin{figure}
\centerline{\epsfig{file=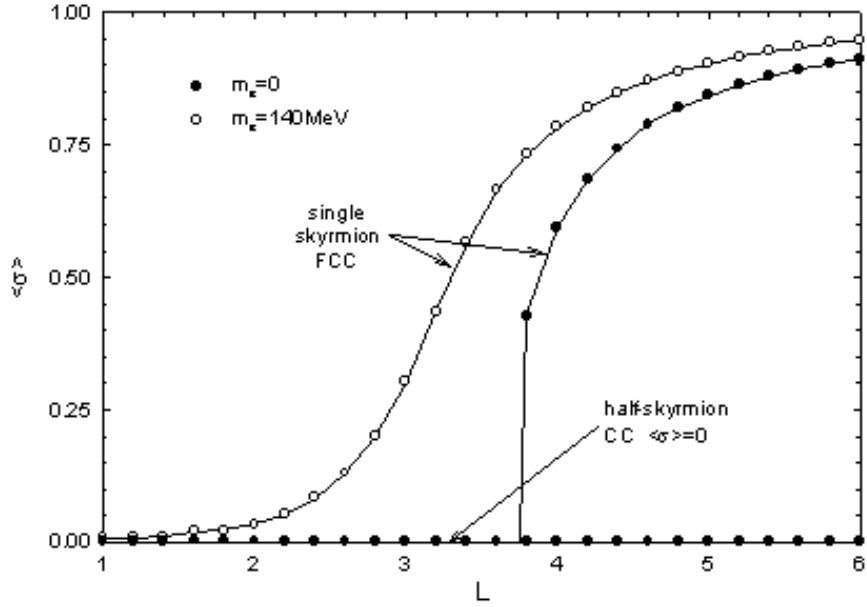,width=12cm,angle=0}}
\caption{$\langle \sigma\rangle$ as a function of the size
parameter $L$. The notation is the same as in Fig.1}
\end{figure}

\begin{figure}
\centerline{\epsfig{file=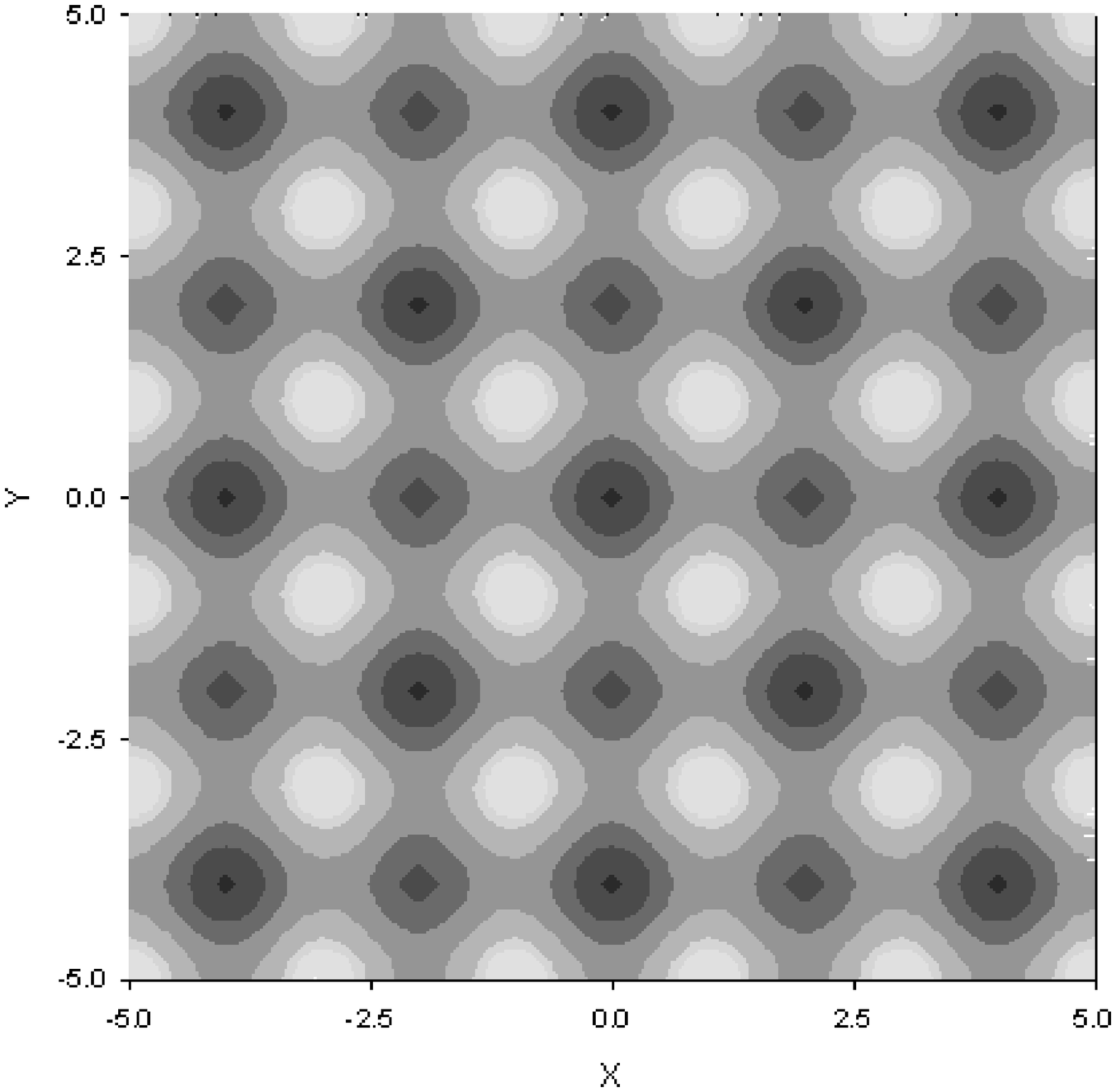,width=8cm,angle=0}
\epsfig{file=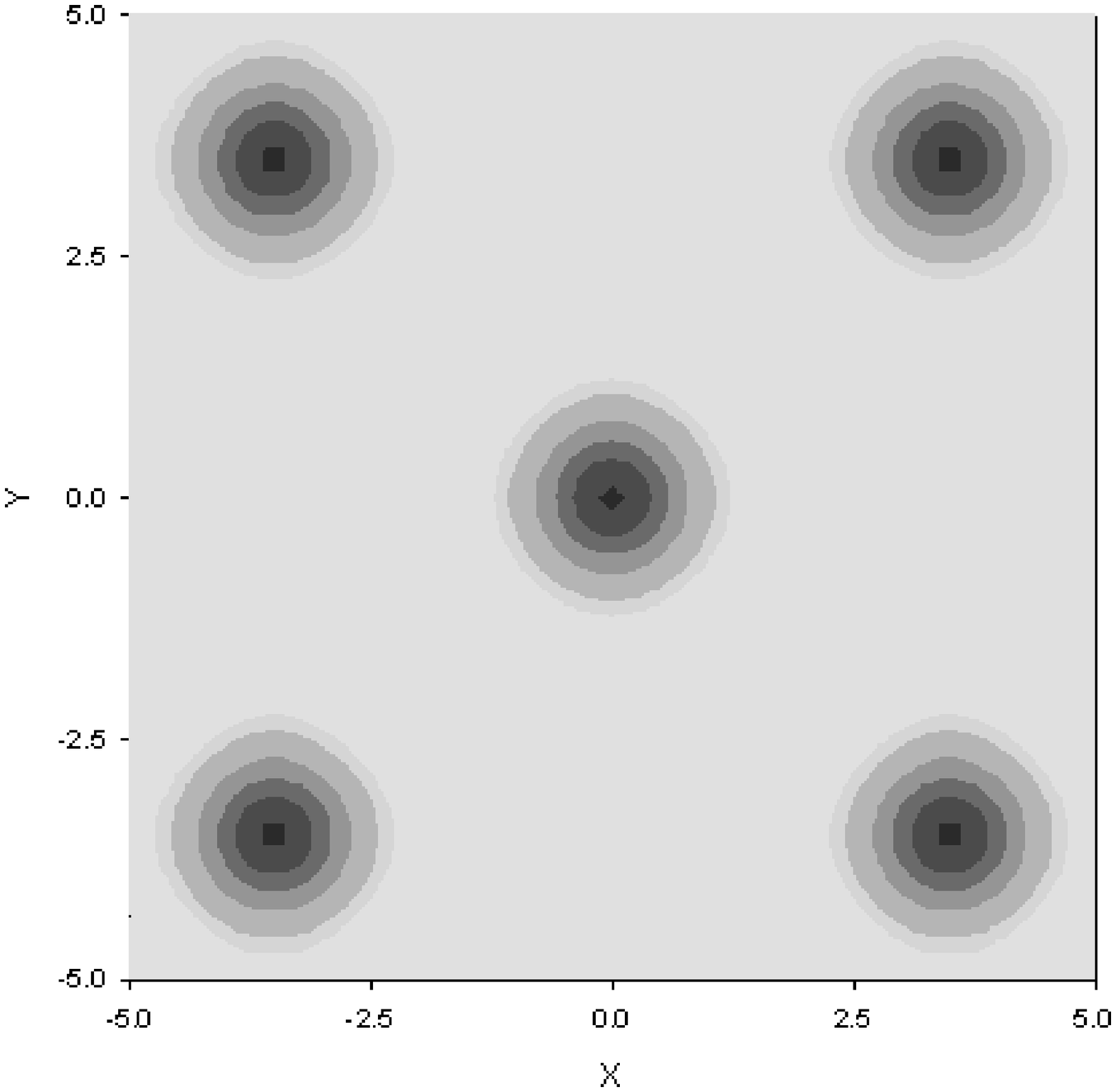,width=8cm,angle=0}}
 \caption{Local baryon
number densities at $L=3.5$ and $L=2.0$ with massive pions. For
$L=2.0$ the system is (almost) a half-skyrmion in a CC crystal
configuration.}

\end{figure}


\section{Pions in Skyrmion Matter}
Let $U_0(\vec{x}) = \sigma + i \vec{\tau}\cdot\vec{\pi}$ be the
static skyrmion crystal configuration obtained so far.  Then, we
incorporate on it fluctuating {\em time-dependent} pion fields
through the ansatz~\cite{saito86}
\begin{equation}
U(\vec{x},t)=\sqrt{U_\pi } U_0 (\vec{x})\sqrt{U_\pi},
\end{equation}
where
\begin{equation}
U_\pi=\exp\bigg(i\vec{\tau}\cdot\vec{\phi}(x)/f_\pi\bigg),
\end{equation}
with $\vec{\phi}$ describing the fluctuating pions. Substituting
this ansatz into the Skyrme Lagrangian (\ref{Lsk}) and expanding
the Lagrangian in terms of pions up to second order, we obtain the
Lagrangian for the pion in the skyrmion matter which is given by
\begin{equation}
{\cal L}={\cal L}_B(U_0) + {\cal L}_M(U_\pi) + {\cal L}_I,
\end{equation}
where ${\cal L}_B(U_0)$ is the Lagrangian density for the skyrmion
matter described by the static configuration $U_0$, ${\cal L}_M$
is for the Lagrangian density for the pions, and finally ${\cal
L}_I$ describes the interaction between the fluctuating pions and
the skyrmion matter. Up to second order in the pion fields, ${\cal
L}_M$ is simply the free pion Lagrangian to
\begin{equation}
{\cal L}_M=\frac{1}{2}\partial _\mu \phi_a \partial^\mu \phi_a
-\frac{1}{2} m_\pi^2 \phi_a^2.
\label{LM}
\end{equation}
${\cal L}_I$ is quite complicated and is given by
\begin{eqnarray}
{\cal L}_I &=& \textstyle \frac{1}{2} \dot{\phi}_a G^{ab}(\vec{x}) \dot{\phi}_b
 - \frac{1}{2} (\partial_i \phi_a) H^{ab}_{ij}(\vec{x}) (\partial_j \phi_b)
\nonumber\\
&&\displaystyle -  \frac{1}{f_\pi}(\partial_i
\phi_a)A^a_i(\vec{x}) -
\frac{1}{2f_\pi^2}\epsilon_{abc}(\partial_i \phi_a)\phi_b
V^c_i(\vec{x})
\nonumber \\
&& +\textstyle \frac{1}{4}m_\pi^2\phi_a^2{\rm Tr}(1-U_0) +\frac{1}{2}m_\pi^2 f_\pi\phi_a{\rm Tr}(i\tau_a U_0)
\label{LI}
\end{eqnarray}
where
\begin{eqnarray}
G^{ab}(\vec{x}) &=& \frac{1}{4} {\rm Tr} \bigg(\tau^a U_0 \tau^b
U_0^\dagger-\tau^a \tau^b\bigg)
+ \frac{1}{32e^2f_\pi^2}{\rm Tr}\bigg(
[R_i,\tilde{\tau}^a] [R_i,\tilde{\tau}^b] \bigg),
\label{G} \\
H^{ab}_{ij}(\vec{x}) &=&  G^{ab}\delta_{ij} +
\frac{1}{32e^2f_\pi^2}{\rm Tr}\bigg( [R_i , R_j][\tilde{\tau}^a,
\tilde{\tau}^b] -[R_i, \tilde{\tau}^b][R_j,\tilde{\tau}^a]\bigg),
\label{H} \\
V^a_i(\vec{x}) &=& \frac{i}{4}f_\pi^2 {\rm Tr}[(L_i + R_i )\tau_a]
 + \frac{i}{16e^2} {\rm Tr} \bigg([L_j, \tau_a][L_i, L_j]
 + [R_j, \tau_a][R_i, R_j] \bigg),
\label{V} \\
A^a_i(\vec{x}) &=& \frac{i}{4}f_\pi^2 {\rm Tr}[(L_i - R_i)\tau_a]
 + \frac{i}{16e^2} {\rm Tr} \bigg([L_j, \tau_a][L_i, L_j]
 - [R_j, \tau_a][R_i, R_j] \bigg)
\label{A}
\end{eqnarray}
with $L_i=(\partial_i U_0^\dagger)U_0$ and
$R_i=(\partial_i U_0)U_0^\dagger$
and $\tilde{\tau}^a=\tau^a + U_0 \tau^a U_0^\dagger$.
The currents $V^a_i$ and $A^a_i$ produced by the static background
field satisfy the following conservation equations:
\begin{eqnarray}
\vec{\nabla}\cdot \vec{V}^a &=& 0, \\
\vec{\nabla}\cdot \vec{A}^a &=&-\frac{1}{2}m_\pi^2 f_\pi^2{\rm Tr}(i\tau_a
U_0).
\end{eqnarray}

The structure of our Lagrangian comes out to be very similar to
that of chiral perturbation theory, eq.~(\ref{ChPT}), and
refs.~\cite{YMK94,TW95}. Let us use this fact to make some crude
estimates.

The interactions of pions and skyrmions as given by eq.~(\ref{LI})
modifies the mass term in eq.~(\ref{LM}) as
\begin{equation}
\textstyle \frac12 m_\pi^2 \sigma(\vec{x}) \phi_a^2.
\end{equation}
This implies that the {\em local} effective pion mass can be even
imaginary in the central part of the skyrmions where $\sigma$ is
negative. By taking the average over space of $\sigma$, we obtain
an effective mass of the pion in the baryonic medium as
\begin{equation}
m^*_\pi \sim m_\pi \langle \sigma \rangle^{1/2}.
\label{mpi0}\end{equation}

The kinetic term is also modified by the background, with respect
to the free case, with an additional factor $G^{ab}(\vec{x})$
given by
\begin{eqnarray}
G^{ab}(\vec{x})&=& -(\pi^2 \delta_{ab} - \pi_a \pi_b)
+\frac{1}{e^2 f_\pi^2} \left\{ \bigg( (\partial_i \sigma )^2 +
(\partial_i \vec\pi )^2 \bigg) \bigg(\sigma^2 \delta_{ab} +
\pi_a\pi_b\bigg) \right.
\\
&& \left. \hskip 10em - \bigg(\partial_i \sigma \pi_a - \sigma
\partial_i \pi_a \bigg) \bigg(\partial_i \sigma \pi_b - \sigma
\partial_i \pi_b \bigg) \right\}.
\end{eqnarray}
The first term in $G^{ab}(\vec{x})$, $
-\pi^2\delta_{ab}+\pi_a\pi_b$, arises from the quadratic term of
the Skyrme model Lagrangian, eq.~(\ref{Lsk}), and corresponds to
the term containing $D^{00}$ in eq.~(\ref{ChPT}). Let us denote it
by $\Gamma^{ab}(\vec{x})$. It changes the kinetic term in
eq.~(\ref{LM}) to
\begin{equation}
\textstyle \frac12 f_\pi^2 (1+ \Gamma^{ab}(\vec{x})) \dot{\phi}_a
\dot{\phi}_b. \label{D00ab}\end{equation}
Ignoring that the inverse of this matrix $\Gamma^{ab}$ becomes
singular at the points where $\pi^2=1$ and taking the diagonal
element, for example, $\Gamma^{11}$, eq.~(\ref{D00ab}) implies that
the pion decay constant behaves {\em locally} as
\begin{equation}
f_\pi^*(\vec{x}) \sim f_\pi (1+\Gamma^{11}(\vec{x}))^{\frac12}.
\end{equation}
The average over the space leads to an effective pion decay
constant in the medium $f^*_\pi$ given by~\footnote{Comparing
eqs.~(\ref{LI}), (\ref{G}) and (\ref{H}) with eq.~(\ref{ChPT}) it
is apparent that to second order  ${\cal L}_I$ leads to different
time and space components for $f_\pi^*$. We keep the present
discussion to leading order where they are equal.}
\begin{equation}
f^*_\pi \sim f_\pi (1-\textstyle \frac23 \langle \pi^2
\rangle)^{\frac12}. \label{fpi1}
\end{equation}
If we absorb this change of $f_\pi$ in the fields to keep the
functional form of the theory as in the free case, eq.~(\ref{mpi0})
is modified to
\begin{equation}
m^*_\pi \sim m_\pi \left( \frac{\langle \sigma \rangle}{1-\frac23
\langle \pi^2\rangle} \right)^\frac12.
\label{mpi1}\end{equation}
This equation shows the direct relation between the pion mass in
the medium, $m^*$, and the parameter $\langle\sigma\rangle$ which
signalled the phase transition in the case of skyrmion matter. We
next show that chiral symmetry is restored in dense matter.

In Fig.~4 we show the estimates of $f^*_\pi/f_\pi$ and
$m^*_\pi/m_\pi$ as a function of the density. As the density
increases, $f^*_\pi$ decreases to $\sim 0.65 f_\pi$ and then it
remains constant at that value \footnote{These estimates are based
on the lowest order approximation and should be valid only for low
densities.}. Note that $\langle \sigma\rangle^\frac12$ has the
same slope at low densities, which leads to $m^*_\pi/m_\pi \sim 1$
at low densities. Since at higher densities $\langle \pi^2\rangle$
becomes a constant, $m^*_\pi/m_\pi$ decreases like $\langle
\sigma\rangle^{1/2}$ with a factor which is greater than 1. As the
density increases, higher order terms in $\rho$ come to play
important roles and $m^*_\pi/m_\pi$ decreases.

The slope of $\langle \sigma\rangle$ at low density is
approximately 1/3. If we expand $\langle\sigma\rangle$ about
$\rho=0$ and compare it with eq.~(\ref{ChPT}), we obtain
\begin{equation}
\langle\sigma\rangle \sim 1 - \frac{1}{3} \frac{\rho}{\rho_0} +
\cdots \sim 1 -\frac{\Sigma_{\pi N}}{f_\pi^2 m_\pi^2} \rho +
\cdots,
\end{equation}
which yields $\Sigma_{\pi N} \sim m^2_\pi f_\pi^2 /(3\rho_0) \sim
42$ MeV, which is comparable with the experimental value 45
MeV~\footnote{While this value is widely quoted, there is at
present a considerable controversy on the precise value of this
sigma term. In fact it can even be considerably higher than this.
See \cite{gibbs} for a recent discussion.}. This comparison is
fully justified from the point of view of the $\frac{1}{N}$
expansion since both approaches should produce the same result to
leading order in this expansion. The liner term is ${\cal{O}}$(1).

The length scale is strongly dependent on our choice of the
parameters $f_\pi$ and $e$. Thus one should be aware that the
$\rho$ scale  in Fig.~4 could change quantitatively considerably if
one chooses another parameter set, however the qualitative
behavior will remain unchanged.

\begin{figure}
\centerline{\epsfig{file=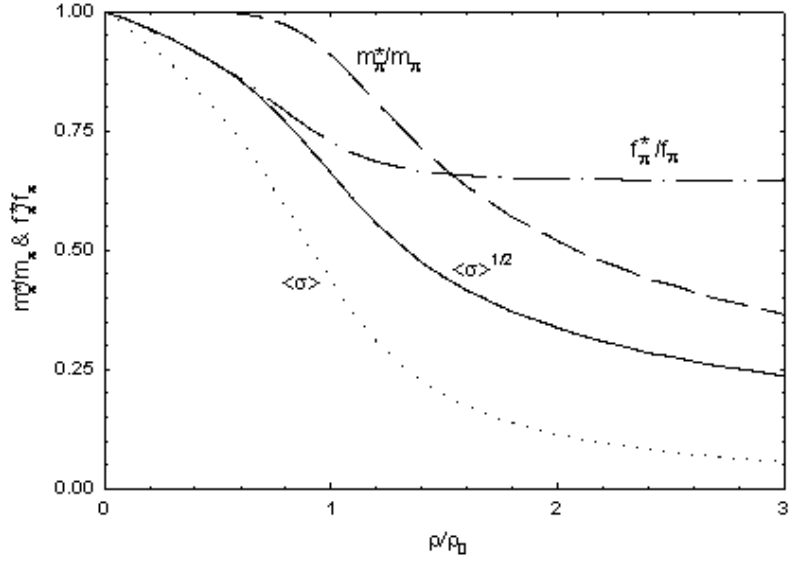,width=12cm,angle=0}}
\caption{Naive estimates of $m^*_\pi/m_\pi$ and $f^*_\pi/f_\pi$ as
functions of the baryon number density of skyrmion matter.}
\end{figure}

We next proceed to a more rigorous derivation of these quantities
using perturbation theory. The presence of the zero modes,
associated with the collective motion of the skyrmion, and the
complexity of $G^{ab}$  make the quantization procedure
non-trivial. Furthermore, $G^{ab}$ becomes singular at the points
where $\sigma(\vec{x})=0$.

Let us obviate the complications associated with the zero modes
and proceed. The momenta conjugate to the pion field $\phi_a$ are
given by
\begin{equation}
\Pi_a = (\delta_{ab} + G^{ab}(\vec{x})) \dot{\phi}_b
\label{P}\end{equation}
which leads to the following Hamiltonian
\begin{equation}
{\cal H} = \Pi_a \dot{\phi}_a -{\cal L}_M - {\cal L}_I.
\end{equation}
It is convenient, as in chiral perturbation
theory~\cite{Weinberg61}, to decompose it into the free pion
Hamiltonian, ${\cal H}_0$, plus the interacting Hamiltonian,
${\cal H}_{I}$, defined by
\begin{eqnarray}
{\cal H}_0 &=& \textstyle\frac12 \dot{\phi}_a \dot{\phi}_a
+\frac12 (\vec{\nabla}\phi_a) \cdot (\vec{\nabla}\phi_a) +\frac12
m_\pi^2 \phi_a \phi_a,
\label{H0} \\
{\cal H}_I &=& \textstyle \frac12 \dot{\phi}_a G^{ab}(\vec{x})
\dot{\phi}_b
 + \frac{1}{2} (\partial_i \phi_a)
 H^{ab}_{ij}(\vec{x}) (\partial_j \phi_b)\nonumber \\
 &&- \frac{1}{2f_\pi^2}\epsilon_{abc}(\partial_i \phi_a)
  \phi_b  V^c_i(\vec{x})
 +\textstyle \frac{1}{2}m_\pi^2\phi_a (\sigma(\vec{x})-1) \phi_a .
\label{HI}
\end{eqnarray}
Note that the momenta $\Pi_a$, defined by eq.~(\ref{P}), and used
in the derivation of the Hamiltonian do not appear in ${\cal
H}_0$. This simplifies matters considerably since only
$\dot{\phi}_a$ will define the canonical momenta of the
quantization procedure. ${\cal H}_0$ defines the free pion
propagator and ${\cal H}_I$ the interaction potentials as
summarized in Fig.~5.

\begin{figure}
\begin{center}
\setlength{\unitlength}{1mm}
\begin{picture}(150,100)
\linethickness{0.8pt}
\put(75,95){\makebox(0,0)[c]{(a)}} \put(58,85){\circle*{1}}
\multiput(58,85)(5,0){7}{\line(1,0){4}} \put(92,85){\circle*{1}}
\put(75,80){\makebox(0,0)[c]{$(p_0,\vec{p})$}}
\put(75,70){\makebox(0,0)[c]{$\displaystyle\frac{1}{p_0^2-\vec{p}^2
- m_\pi^2}$}}
\put(35,75){\makebox(0,0)[c]{(b)}}
\put(35,60){\makebox(0,0)[c]{$\hat{S}(\vec{\ell})$}}
\put(27,62){\makebox(0,0)[b]{a}}
\put(27,59){\makebox(0,0)[tr]{$(q_0,\vec{q})$}}
\put(43,62){\makebox(0,0)[b]{a}}
\put(43,59){\makebox(0,0)[tl]{$(p_0,\vec{p})$}}
\put(28,60){\circle*{1}} \put(35,60){\circle{14}}
\put(42,60){\circle*{1}}
\put(35,45){\makebox(0,0)[c]{$\displaystyle m_\pi^2
\left(\sum_{\vec{\ell}}\hat{S}(\vec{\ell})\delta^3
(\vec{p}-\vec{q}-\vec{\ell})-\delta^3 (\vec{p}-\vec{q})\right)$}}
\put(110,75){\makebox(0,0)[c]{(c)}}
\put(110,60){\makebox(0,0)[c]{$\hat{G}^{ab}(\vec{\ell})$}}
\put(102,62){\makebox(0,0)[b]{b}}
\put(102,59){\makebox(0,0)[tr]{$(q_0,\vec{q})$}}
\put(118,62){\makebox(0,0)[b]{a}}
\put(118,59){\makebox(0,0)[tl]{$(p_0,\vec{p})$}}
\put(103,60){\circle*{1}} \put(110,60){\circle{14}}
\put(117,60){\circle*{1}}
\put(110,45){\makebox(0,0)[c]{$-\displaystyle p_0^2
\left(\sum_{\vec{\ell}}\hat{G}^{ab}(\vec{\ell})\delta^3
(\vec{p}-\vec{q}-\vec{\ell})\right)$}}
\put(35,35){\makebox(0,0)[c]{(d)}}
\put(35,20){\makebox(0,0)[c]{$\hat{H}^{ab}_{ij}(\vec{\ell})$}}
\put(27,22){\makebox(0,0)[b]{b}}
\put(27,19){\makebox(0,0)[tr]{$(q_0,\vec{q})$}}
\put(43,22){\makebox(0,0)[b]{a}}
\put(43,19){\makebox(0,0)[tl]{$(p_0,\vec{p})$}}
\put(28,20){\circle*{1}} \put(35,20){\circle{14}}
\put(42,20){\circle*{1}}
\put(35,5){\makebox(0,0)[c]{$\displaystyle -p_i p_j
\left(\sum_{\vec{\ell}}\hat{H}^{ab}_{ij}(\vec{\ell})\delta^3
(\vec{p}-\vec{q}-\vec{\ell})\right)$}}
\put(110,35){\makebox(0,0)[c]{(e)}}
\put(110,20){\makebox(0,0)[c]{$\hat{V}^{c}(\vec{\ell})$}}
\put(102,22){\makebox(0,0)[b]{b}}
\put(102,19){\makebox(0,0)[tr]{$(q_0,\vec{q})$}}
\put(118,22){\makebox(0,0)[b]{a}}
\put(118,19){\makebox(0,0)[tl]{$(p_0,\vec{p})$}}
\put(103,20){\circle*{1}} \put(110,20){\circle{14}}
\put(117,20){\circle*{1}}
\put(110,5){\makebox(0,0)[c]{$-\displaystyle i \epsilon_{abc}
(p+q)_i\left(\sum_{\vec{\ell}}\hat{V}^{c}_i(\vec{\ell})\delta^3
(\vec{p}-\vec{q}-\vec{\ell})\right)$}}

\end{picture}
\end{center}
\caption{Free propagator and interactions between the pion fields
and background skyrmion matter.}
\end{figure}

We define the effective mass of the pion in skyrmion matter as the
pole in the pion propagator in the limit of $\vec{p} \rightarrow
0$,
\begin{equation}
G(p) = \frac{1}{p^2 - m_\pi^2 - \Sigma(p)},
\end{equation}
where $\Sigma(p)$ is the pion self-energy in matter. Thus, the
interactions associated with the spatial derivatives acting on the
pion fields will not play any role in the process up to second
order in the interactions. Only the interactions characterized by
$G^{ab}$ and $\sigma$ will be active to this order.

Due to the crystal structure of the background fields in the
homogeneous phase, $G^{ab}(\vec{x})$ and $\sigma(\vec{x})$ can be
expressed as
\begin{eqnarray}
\sigma(\vec{x}) &=& \sum_{\ell_x,\ell_y,\ell_z} \hat{S}({\vec{\ell}})
\ \exp(i\vec{\ell}\cdot\vec{x} \pi /L),
\label{normalized_sigma}\\
G^{ab}(\vec{x}) &=& \sum_{\ell_x,\ell_y,\ell_z}
\hat{G}^{ab}({\vec{\ell})}
{} \exp(i\vec{\ell}\cdot\vec{x} \pi /L),
\label{Gab}
\end{eqnarray}
with the help of a set of integers
$\vec{\ell}=(\ell_x,\ell_y,\ell_z)$. Thus, the corresponding
$\hat{\sigma}(\vec{p})$ and $\hat{G}^{ab}(\vec{p})$ in  momentum
space  have the following structure,
\begin{equation}
\hat{\sigma}(\vec{p}) =\sum_{\vec{\ell}} \hat{S}({\vec{\ell}}) \
\delta^3 (\vec{p}-\vec{\ell}), \hskip 3em
\hat{\gamma}^{ab}(\vec{p}) =\sum_{\vec{\ell}}
\hat{G}^{ab}({\vec{\ell}}) {} \delta^3 (\vec{p}-\vec{\ell}).
\end{equation}
The FCC crystal yields non-vanishing expansion coefficients only
for the integer set listed in Tab.~2. Furthermore, the
coefficients have the following symmetries
\begin{eqnarray}
\hat{S}({\ell_x,\ell_y,\ell_z})
=+\hat{S}({-\ell_x,\ell_y,\ell_z})
=+\hat{S}({\ell_x,-\ell_y,\ell_z})
=+\hat{S}({\ell_x,\ell_y,-\ell_z})
\nonumber\\
\hat{G}^{11}({\ell_x,\ell_y,\ell_z})
=\hat{G}^{11}({-\ell_x,\ell_y,\ell_z})
=+\hat{G}^{11}({\ell_x,-\ell_y,\ell_z})
=+\hat{G}^{11}({\ell_x,\ell_y,-\ell_z})
\\
\hat{G}^{12}({\ell_x,\ell_y,\ell_z})
=-\hat{G}^{12}({-\ell_x,\ell_y,\ell_z})
=-\hat{G}^{12}({\ell_x,-\ell_y,\ell_z})
=+\hat{G}^{12}({\ell_x,\ell_y,-\ell_z})
\nonumber
\end{eqnarray}

\begin{table}
\caption{Possible integer values for the expansion coefficients
$\beta_{\vec{\ell}}$, $\gamma^{11}_{\vec{\ell}}$ and
$\gamma^{12}_{\vec{\ell}}$.}
\begin{center}
\begin{tabular}{cccc|cccc|cccc}
\hline
$\beta_{\vec{\ell}}$       & $\ell_x$ & $\ell_y$ & $\ell_z$ &
$\gamma^{11}_{\vec{\ell}}$ & $\ell_x$ & $\ell_y$ & $\ell_z$ &
$\gamma^{12}_{\vec{\ell}}$ & $\ell_x$ & $\ell_y$ & $\ell_z$ \\
\hline
(i) & even & even & even & (i) & odd & odd & odd & (i) & odd & odd & even \\
(ii) & odd & odd & odd & (ii) & even & even & even & (ii) & even & even & odd \\
\hline
\end{tabular}
\end{center}
\end{table}

To the first order in the interactions, the self-energy gets
contributions from $\sigma(\vec{0})$ and $G^{11}(\vec{0})$ given
by
\begin{equation}
\Sigma^{I}(p_0,\vec{p}\rightarrow 0)
 = m_\pi^2 (\hat{S}(\vec{0}) -1 ) - p_0^2 \hat{G}^{11}(\vec{0}).
\label{Z}\end{equation}
To this order, the effective pion mass $m_\pi^*$ becomes
\begin{equation}
m_\pi^{*I}  = \left(
\frac{\hat{S}(\vec{0})}{1+\hat{G}^{11}(\vec{0})}
\right)^{1/2} m_\pi.
\label{mpiI}\end{equation}
Since $\hat{S}(\vec{0})$ is nothing but $\langle \sigma \rangle$,
if we ignore the denominator, we can interpret our result as the
naive estimate (\ref{mpi0}). Furthermore, keeping  only the first
order terms in $G^{11}(\vec{x})$, the result coincides with our
estimate (\ref{mpi1}). Thus, eq.~(\ref{mpiI}) differs from
eq.~(\ref{mpi1}) by the presence of the higher order term coming
from the Skyrme term. Since the latter is positive definite,
$\hat{G}^{11}(\vec{0})$ is always larger than $-\frac23 \langle
\pi^2\rangle$ appearing in eq.~(\ref{mpi1}). According to our
numerical results, however, $\hat{G}^{11}(\vec{0})$ takes negative
values for a wide range of $\rho$ down to the density where $E/B$
has the minimum. Thus, eq.~(\ref{mpiI}) tells us that
${m_\pi^*}/{m_\pi}$ decreases slower than $\langle\sigma
\rangle^{1/2}$ in this range of $\rho$, but faster than the naive
estimate (\ref{mpi1}).

If we include the second order diagrams as shown in Fig.~6, we
obtain the self-energy as
\begin{eqnarray}
\Sigma^{II}(p_0,\vec{0}) &=& \Sigma^{I}(p_0,\vec{0}) \nonumber\\
&&+\sum_{\vec{\ell} \neq \vec{0} } \left(
\frac{\hat{S}(\vec{\ell})\hat{S}(-\vec{\ell}) m_\pi^4}
{p_0^2 -m_\pi^2 - \ell^2 \pi^2/L^2}
 - \frac{2\hat{S}(\vec{\ell})\hat{G}^{11}(-\vec{\ell}) m_\pi^2 p_0^2}
{p_0^2 -m_\pi^2 - \ell^2 \pi^2/L^2} \right. \nonumber\\
&& \left.
+\frac{(\hat{G}^{11}(\vec{\ell})\hat{G}^{11}(-\vec{\ell}) + 2
\hat{G}^{12}(\vec{\ell})\hat{G}^{21}(-\vec{\ell})\ )p_0^4 } {p_0^2
-m_\pi^2 - \ell^2 \pi^2/L^2} \right). \label{SII}\end{eqnarray}

\begin{figure}
\begin{center}
\setlength{\unitlength}{1mm}
\begin{picture}(150,60)(0,40)
\linethickness{0.8pt}
\put(8,92){\makebox(0,0)[b]{\scriptsize 1}}
\put(8,89){\makebox(0,0)[tr]{\scriptsize $(p_0,\vec{0})$}}
\put(22,92){\makebox(0,0)[b]{\scriptsize 1}}
\put(22,89){\makebox(0,0)[tl]{\scriptsize $(p_0,\vec{0})$}}
\put(9,90){\circle*{1}} \put(15,90){\circle{11}}
\put(15,90){\makebox(0,0)[c]{$\Sigma$}} \put(21,90){\circle*{1}}
\put(35,90){\makebox(0,0)[c]{=}}
\put(48,92){\makebox(0,0)[b]{\scriptsize 1}}
\put(48,89){\makebox(0,0)[tr]{\scriptsize $(p_0,\vec{0})$}}
\put(62,92){\makebox(0,0)[b]{\scriptsize 1}}
\put(62,89){\makebox(0,0)[tl]{\scriptsize $(p_0,\vec{0})$}}
\put(49,90){\circle*{1}} \put(55,90){\circle{11}}
\put(55,90){\makebox(0,0)[c]{G$^{11}$}} \put(61,90){\circle*{1}}
\put(75,90){\makebox(0,0)[c]{+}}
\put(88,92){\makebox(0,0)[b]{\scriptsize 1}}
\put(88,89){\makebox(0,0)[tr]{\scriptsize $(p_0,\vec{0})$}}
\put(102,92){\makebox(0,0)[b]{\scriptsize 1}}
\put(102,89){\makebox(0,0)[tl]{\scriptsize $(p_0,\vec{0})$}}
\put(89,90){\circle*{1}} \put(95,90){\circle{11}}
\put(95,90){\makebox(0,0)[c]{S}} \put(101,90){\circle*{1}}
\put(35,70){\makebox(0,0)[c]{+}}
\put(48,72){\makebox(0,0)[b]{\scriptsize 1}}
\put(48,69){\makebox(0,0)[tr]{\scriptsize $(p_0,\vec{0})$}}
\put(62,72){\makebox(0,0)[b]{\scriptsize b}}
\put(49,70){\circle*{1}} \put(55,70){\circle{11}}
\put(55,70){\makebox(0,0)[c]{G$^{1b}$}} \put(61,70){\circle*{1}}
\multiput(61,70)(3,0){4}{\line(1,0){2}}
\put(66,65){\makebox(0,0)[c]{$(p_0,\vec{\ell})$}}
\put(71,72){\makebox(0,0)[b]{\scriptsize b}}
\put(85,72){\makebox(0,0)[b]{\scriptsize 1}}
\put(72,70){\circle*{1}} \put(78,70){\circle{11}}
\put(78,70){\makebox(0,0)[c]{G$^{b1}$}} \put(84,70){\circle*{1}}
\put(85,69){\makebox(0,0)[tl]{\scriptsize $(p_0,\vec{0})$}}
\put(95,70){\makebox(0,0)[c]{+}}
\put(103,72){\makebox(0,0)[b]{\scriptsize 1}}
\put(104,69){\makebox(0,0)[tr]{\scriptsize $(p_0,\vec{0})$}}
\put(117,72){\makebox(0,0)[b]{\scriptsize 1}}
\put(104,70){\circle*{1}} \put(110,70){\circle{11}}
\put(110,70){\makebox(0,0)[c]{S}} \put(116,70){\circle*{1}}
\multiput(116,70)(3,0){4}{\line(1,0){2}}
\put(121,65){\makebox(0,0)[c]{$(p_0,\vec{\ell})$}}
\put(126,72){\makebox(0,0)[b]{\scriptsize 1}}
\put(140,72){\makebox(0,0)[b]{\scriptsize 1}}
\put(127,70){\circle*{1}} \put(133,70){\circle{11}}
\put(133,70){\makebox(0,0)[c]{S}} \put(139,70){\circle*{1}}
\put(139,69){\makebox(0,0)[tl]{\scriptsize $(p_0,\vec{0})$}}
\put(35,50){\makebox(0,0)[c]{+}}
\put(48,52){\makebox(0,0)[b]{\scriptsize 1}}
\put(48,49){\makebox(0,0)[tr]{\scriptsize $(p_0,\vec{0})$}}
\put(62,52){\makebox(0,0)[b]{\scriptsize 1}}
\put(49,50){\circle*{1}} \put(55,50){\circle{11}}
\put(55,50){\makebox(0,0)[c]{G$^{11}$}} \put(61,50){\circle*{1}}
\multiput(61,50)(3,0){4}{\line(1,0){2}}
\put(66,45){\makebox(0,0)[c]{$(p_0,\vec{\ell})$}}
\put(71,52){\makebox(0,0)[b]{\scriptsize 1}}
\put(85,52){\makebox(0,0)[b]{\scriptsize 1}}
\put(72,50){\circle*{1}} \put(78,50){\circle{11}}
\put(78,50){\makebox(0,0)[c]{S}} \put(84,50){\circle*{1}}
\put(85,49){\makebox(0,0)[tl]{\scriptsize $(p_0,\vec{0})$}}
\put(95,50){\makebox(0,0)[c]{+}}
\put(103,52){\makebox(0,0)[b]{\scriptsize 1}}
\put(104,49){\makebox(0,0)[tr]{\scriptsize $(p_0,\vec{0})$}}
\put(117,52){\makebox(0,0)[b]{\scriptsize 1}}
\put(104,50){\circle*{1}} \put(110,50){\circle{11}}
\put(110,50){\makebox(0,0)[c]{S}} \put(116,50){\circle*{1}}
\multiput(116,50)(3,0){4}{\line(1,0){2}}
\put(121,45){\makebox(0,0)[c]{$(p_0,\vec{\ell})$}}
\put(126,52){\makebox(0,0)[b]{\scriptsize 1}}
\put(140,52){\makebox(0,0)[b]{\scriptsize 1}}
\put(127,50){\circle*{1}} \put(133,50){\circle{11}}
\put(133,50){\makebox(0,0)[c]{G$^{11}$}} \put(139,50){\circle*{1}}
\put(139,49){\makebox(0,0)[tl]{\scriptsize $(p_0,\vec{0})$}}

\end{picture}
\end{center}
\caption{Diagrams used to evaluate the self-energy up to second
order in the interactions. Here, $b$ runs over 1,2,3 and
$\vec{\ell} \neq 0$.}
\end{figure}

The resulting effective pion mass can be found numerically by
solving the equation
\begin{equation}
p_0^2 - m_\pi^2 - \Sigma^{II}(p_0,\vec{0}) = 0,
\label{pole}\end{equation}
after substituting $p_0 \equiv m_\pi^{*II}$.

In the chiral limit where $m_\pi=0$, eq.~(\ref{pole}) always
provides us with the trivial solution  $p_0 = 0$, which supports
the consistency of our process. We show in Fig.~7 the numerical
results on $m_\pi^* / m_\pi$ as a function of the density $\rho$.
We have taken in eq.~(\ref{SII}) up to $|\vec{\ell}|^2<6$ Fourier
components. Although, there is no apparent reason for the
interactions to be weak, $m^{*II}_\pi$ is very close to
$m^{*I}_\pi$. The Fourier expansion for $\hat{S}(\vec{p})$ and
$\hat{G}^{ab}(\vec{p})$ can only be done for the homogeneous
skyrmion crystal.

\begin{figure}
\begin{center}
\centerline{\epsfig{file=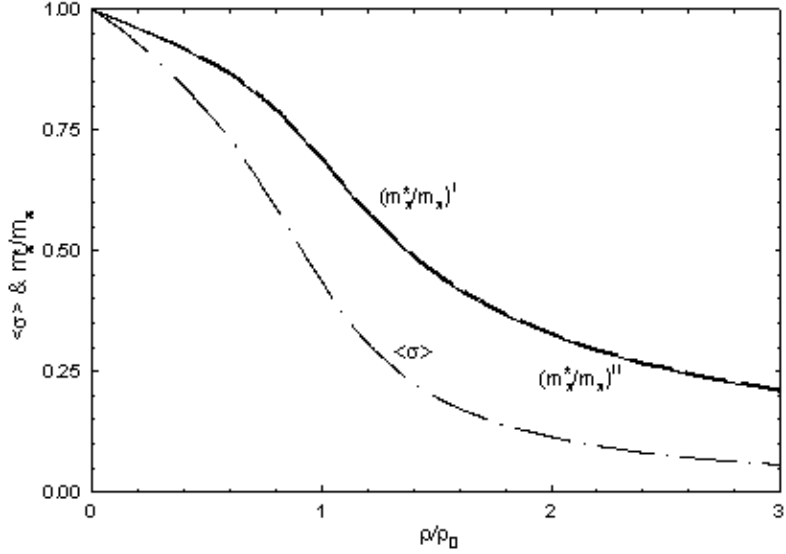,width=12cm,angle=0}}
\end{center}
\caption{$m^{*I}_\pi$ and $m^{*II}_\pi$ as a function of the
density. We have used $f_\pi$=93MeV and $e=4.75$ in the
calculation.}
\end{figure}

Now that we have developed the perturbative expansion, we may
check our estimate eq.~(\ref{fpi1}) of the pion decay constant in
the medium. We proceed by calculating it directly from the
appropriate expectation value of the axial current.

The total axial current of the system can be written as
\begin{eqnarray}
{\cal A}_{\mu, a}&=&f_\pi\partial_\mu\phi_a(x) +f_\pi G^{ab}(\vec{x})
\partial_\mu\phi_b(x)+f_\pi H^{ab}_{\mu\nu}(\vec{x})
\partial^\nu\phi_b(x)\nonumber\\
&&+\frac{1}{2f_\pi}\epsilon_{abc}\phi_b(x)V_{\mu,c}(\vec{x})
+A_{\mu,a}(\vec{x}) ,\label{TAC}
\end{eqnarray}
where $A_{\mu,a}$ and $V_{\mu,a}$ are the axial and vector
currents from the background skyrmion matter, respectively. The
total axial current satisfies the following conservation relation,
\begin{equation}
\partial^\mu {\cal A}_{\mu,a}=-f_\pi m_\pi^2\sigma(\vec{x})
\phi_a(x)-\frac{1}{2f_\pi}\epsilon_{abc}\partial^\mu
\phi_b(x)V_{\mu,c}(\vec{x})+\frac{f_\pi^2}{2}m_\pi^2{\rm
Tr}(i\tau_a U_0).
\end{equation}
The axial current of the fluctuating pion is obtained from
eq.~(\ref{TAC}) by omitting the background (last) term of the
equation. Let us use the symbol $A_{\mu,a}^\phi$ for the in medium
pion axial current. Its conservation  leads to the equation of
motion,
\begin{equation}
\partial^\mu A_{\mu,a}^\phi = -f_\pi m_\pi^2\sigma(\vec{x})
\phi_a(x)-\frac{1}{2f_\pi}\epsilon_{abc}\partial^\mu
\phi_b(x)V_{\mu,c}(\vec{x}).
\end{equation}
Applying the LSZ reduction formalism, the appropriate matrix
element of the pion axial current in the ground state of skyrmion
matter is,
\begin{equation}
\langle \tilde{0}|A_{\mu,1}^\phi (x)|\pi_1(q) \;
\tilde{0}\rangle=-i\sqrt{Z}f_\pi q_\mu(1+G^{11}(\vec{x}))
e^{-iq\cdot x},
\end{equation}
where we have used $\pi^1$ to perform the calculation, exploiting
isospin symmetry, and $Z$ is the wave function renormalization
constant obtained as $Z^{-1}=(1+G^{11}(\vec{0}))$ in
eq.~(\ref{Z}). The ground state of skyrmion matter is denoted by
$|\tilde{0}\rangle$. In the spirit of our approximation we only
consider the $\mu=0$ component, to which $H^{ab}_{\mu\nu}$ does
not contribute. If we regard the quantity appearing in the right
hand side of the above equation as the in-medium pion decay
constant and take the spatial average of $G^{11}(\vec{x})$, we
obtain,
\begin{equation}
f_\pi^*=f_\pi(\sqrt{Z})^{-1}=f_\pi(1+G^{11}(\vec{0}))^{1/2},
\label{fp}
\end{equation}
which supports eq.~(\ref{fpi1}). Similarly, the matrix element for
the conservation relation becomes
\begin{equation}
\langle \tilde{0}|\partial^\mu A_{\mu,1}^\phi(x)|\pi_1(q)
\;\tilde{0}\rangle= -\sqrt{Z}f_\pi m_\pi^2\sigma(\vec{x}).
\end{equation}
Again, if we take the quantity in the right hand as
$-f_\pi^{*2}m_\pi^{*2}$, this equation gives the following
relation between free and in-medium quantities,
\begin{equation}
\frac{m_\pi^{*2}}{m_\pi^2}\cdot\frac{f_\pi^*}{f_\pi}=\sqrt{Z}
\sigma(\vec{0}).
\end{equation}
Incorporating the relation eq.~(\ref{fp}), we obtain
\begin{equation}
\frac{m_\pi^{*2}}{m_\pi^2}\cdot\frac{f_\pi^{*2}}{f_\pi^2}
=\sigma(\vec{0}),
\end{equation}
which is nothing but eq.~(\ref{mpi1}).

Up to now, we have considered only the ``{\em homogeneous}" phase
where the background fields are in a crystal configuration. We
next proceed to describe the ``{\em inhomogeneous phase}". In
Fig.~8, we present the situation schematically; in (a) the
skyrmions are in the homogeneous crystal phase, while in (b) they
condense to form rather dense lumps of matter and empty spaces.
Our crude approximation consists of describing the dense lumps of
matter by the crystal structure at the minimum energy per baryon
ignoring surface effects. In this approximation $E/B$ of the
inhomogeneous phase (b) will be the same as
$(E/B)_{\mbox{\scriptsize min}}$. That is, the inhomogeneous
matter has a lower energy than the homogeneous one and becomes
stable. In Fig.~1, such energy is represented by a horizontal line
starting at the minimum. Our approximation can be interpreted as a
Maxwell construction between the two phases.

\begin{figure}
\begin{center}
\centerline{\epsfig{file=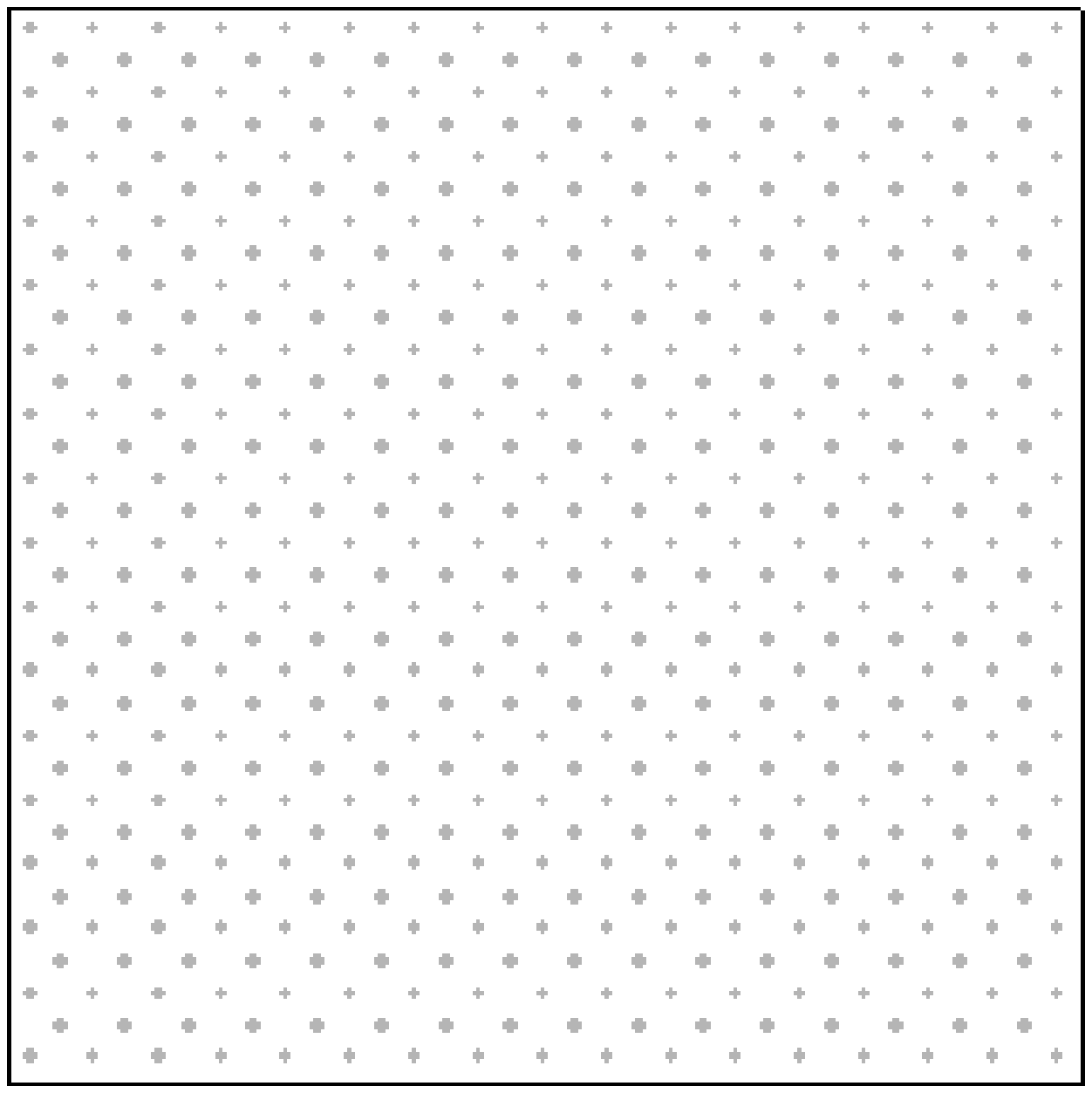,width=8cm,angle=0}
\epsfig{file=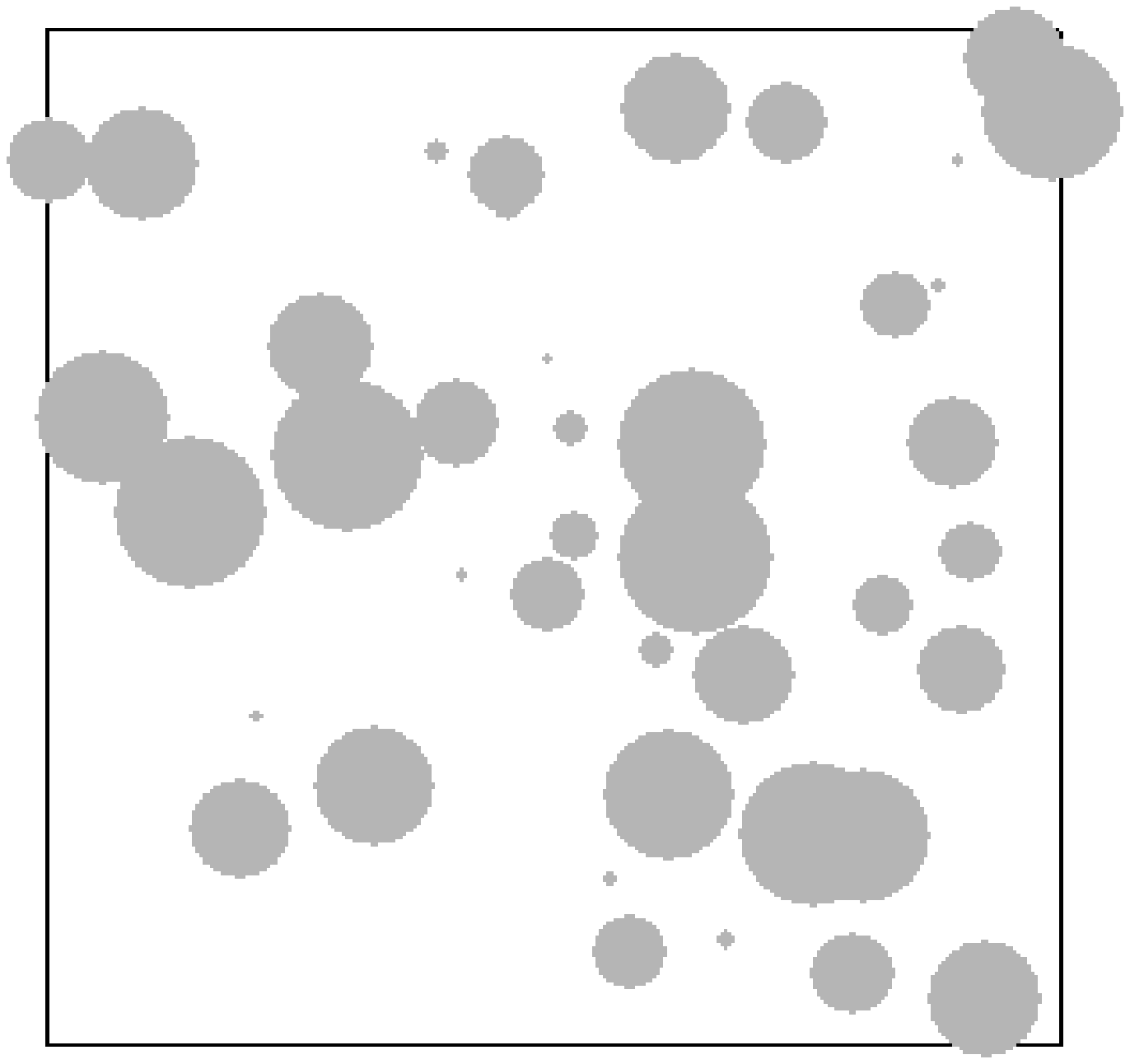,width=8cm,angle=0} }
\end{center}
\caption{(a) The homogeneous but unstable crystal phase and (b)
the inhomogeneous but stable skyrmion matter.}
\end{figure}

In the inhomogeneous matter space is divided into two different
phases. Let us distinguish these  by using the subscripts ``{\em
vac}" denoting the (empty) {\em vacuum} phase and ``{\em min}"
denoting the lumps of skyrmion matter at the {\em minimum} energy.
The expectation values $\langle \sigma \rangle$ can be
approximated by
\begin{equation}
\langle \sigma \rangle^{(i)} \approx \frac{
\langle \sigma \rangle^{(h)}_{\mbox{\scriptsize{min}}}
V_{\mbox{\scriptsize{min}}} +
\langle \sigma \rangle^{(h)}_{\mbox{\scriptsize{vac}}}
V_{\mbox{\scriptsize{vac}}} }
{V_{\mbox{\scriptsize{min}}} + V_{\mbox{\scriptsize{vac}}} }
\label{S_i}\end{equation}
where the superscript ``h" and ``i" denote that the quantities
are for the homogeneous and inhomogeneous phase, respectively.

In the inhomogeneous matter, the total volume of the dense
droplets $V_{\mbox{\scriptsize{min}}}$ is nothing but
$L^3_{\mbox{\scriptsize min}}$ and the volume of the whole space
is $V_{\mbox{\scriptsize{min}}}+V_{\mbox{\scriptsize{vac}}}$.
Thus, eq.~(\ref{S_i}) can be simply written as
\begin{equation}
\langle \sigma \rangle^{(i)} \approx
\langle \sigma \rangle^{(h)}_{\mbox{\scriptsize{vac}}}
- (\langle \sigma \rangle^{(h)}_{\mbox{\scriptsize{vac}}}
-\langle \sigma \rangle^{(h)}_{\mbox{\scriptsize{min}}})
(L_{\mbox{\scriptsize{min}}}/L)^3.
\end{equation}

The dash-dot lines in Fig.~9 represent the value of $\langle
\sigma \rangle$ for the inhomogeneous phase obtained by using the
approximate Maxwell construction defined by eq.~(\ref{S_i}). The
change of $\langle \sigma \rangle$ in the chiral limit is
remarkable. Chiral restoration occurs at the point $L=L_{min}$
where $E/B$ has its lowest value through a phase transition from
the inhomogeneous phase to the homogeneous half-skyrmion CC phase
at $L=L_{\mbox{\scriptsize min}}$. The sharp phase transition may
be an artifact of our naive approximation (\ref{S_i}). In the case
of $m_\pi \neq 0$ also a rapid change in  $\langle \sigma \rangle$
occurs in the transition to the stable inhomogeneous phase,
moreover there is almost no difference between the massive and
massless cases.

\begin{figure}
\begin{center}
\centerline{\epsfig{file=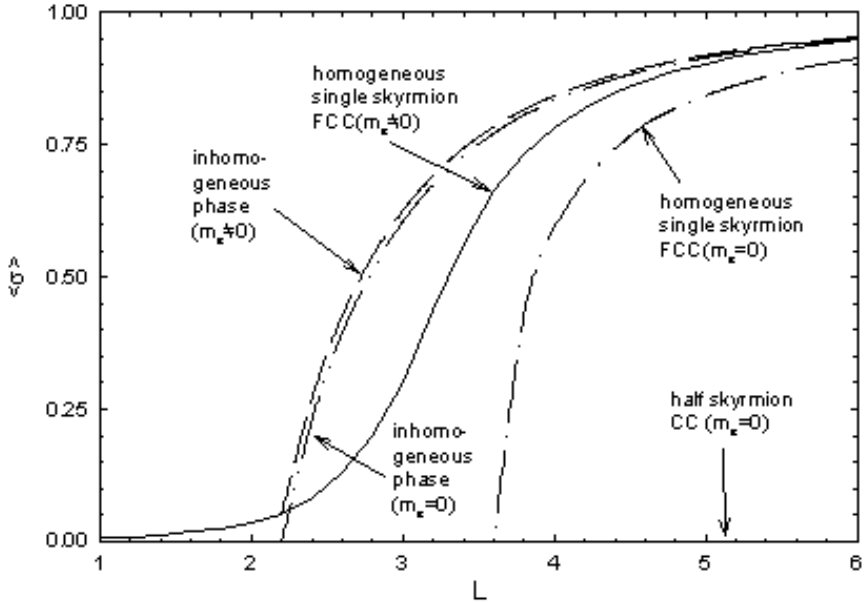,width=12cm,angle=0}}
\end{center}
\caption{Maxwell construction for the homogeneous unstable crystal
phase. We represent the change in $\langle\sigma\rangle$ as a
function of lattice size.}
\end{figure}

We show in Fig.~10  the value of $m^{*i}_\pi$ obtained by
eq.~(\ref{mpi1}) for the inhomogeneous phase.

We learn from our calculation that the dynamics of the pion in the
medium depends not only on the density of the baryonic matter but
also from its detailed structure. It may be possible to interpret
this as simulating the effect of the ``intrinsic" density
dependence required in Harada-Yamawaki
theory~\cite{HY:PR,BR:PR01}.

\begin{figure}
\begin{center}
\centerline{\epsfig{file=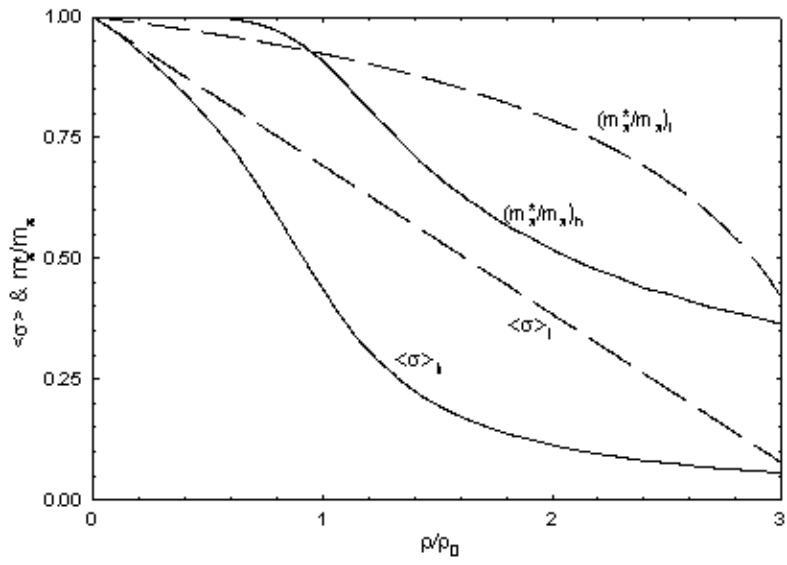,width=12cm,angle=0}}
\end{center}
\caption{Maxwell construction for the homogeneous unstable crystal
phase. We show $m^{*h}_\pi$, $m^{*i}_\pi$,
$\langle\sigma\rangle_h$ and $\langle\sigma\rangle_i$ as a
function of density.}
\end{figure}

\section{Concluding Remarks}
We have studied the dynamics of the pion in a baryonic environment
by incorporating quantum fluctuations to classical skyrmion
matter. The procedure has led to an effective Lagrangian for the
pion in the medium defined in terms of the parameters, $f_\pi^*$
and $m_\pi^*$, which are density-dependent. The fact that we have
a unified hadronic scheme to calculate both baryonic matter and
the interactions of the pion in the background of such matter,
with only a few parameters, allows us to draw important
conclusions. This contrasts with the conventional approach where
the starting point is the matter-free vacuum, with the density
effect taken into account perturbatively.  While our scheme is
still far from being realistic, it contains the right symmetries
and the support of $\frac{1}{N}$ physics, which renders it
qualitatively predictive if not yet quantitatively. We are able to
describe interesting phenomena though we are not able to predict
yet the scales at which they will occur.

The main result of our presentation is that, as the density of
matter increases, chiral symmetry for the dynamics of the pion is
restored, i.e. the effective parameters become those of a chiral
symmetric phase. Two important dynamical mechanisms lead to this
phenomenon. On the one hand the structure of the matter state is
crucial in producing the phase transition. At high densities, the
symmetries of the classical field describing the CC half skyrmion
crystal, conspire to provide the $\sigma$ components with
adequately distributed values on the lattice so that its average
tends to vanish as the crystal becomes more and more precisely
defined. This phenomenon which occurs at the level of the baryonic
state is carried over by the quantum fluctuations to the dynamics
of the pion in this baryonic medium. Thus the important physical
quantities of the pion become density-dependent in terms of the
expectation values of the classical fields and therefore are
essential to this phase transition, which tends to restore chiral
symmetry in the dynamics of the pion. On the other hand our
procedure produces a density dependence which is non-trivial,
i.e., the classical field leads to higher orders in the density
dependence of the parameters. This intricate density dependence is
analogous, if not entirely equivalent, to the intrinsic density
dependence found in hidden local symmetry theory with the vector
manifestation fixed point~\cite{BR:PR01}. The results show that
these additional terms, which do not appear in more conventional
calculations, play an important role in the phase transition,
suggesting that one should not trust results of this phenomenon
obtained with the linear approximation or in few-order low-density
expansions.

It is important to stress that the results just exposed have
required the development of a non-trivial perturbative procedure
which has been solved beyond the leading order in the case of the
effective pion mass.

Ours is not yet a realistic calculation. The model we have used
describes skyrmion matter, not nuclear matter. For instance we do
not have the liquid structure of normal nuclear matter. In fact it
appears nontrivial to arrive at a Fermi liquid from the crystal
structure. Specifically while the in-medium pion decay constant
going proportional to $\langle \sigma\rangle^{1/2}$ up to near
nuclear matter density is consistent with what one expects in the
effective field theory with vector manifestation~\cite{BR:PR01},
its saturation at higher density and non-vanishing as one
approaches the chiral transition indicate that we may be
describing a phase which is not the standard Wigner-Weyl symmetry.
It may be in a pseudo-gap phase where the gap is non-zero though
chiral symmetry is restored, resembling what might be happening in
the ``normal phase" of high $T_c$
superconductivity~\cite{hightc}~\footnote{We are grateful to Kurt
Langfeld for bringing our attention to this reference.}. This may
account for the fact that the in-medium pion mass $decreases$ as
one goes beyond normal matter density, which is at variance with
the results of other approaches.

It should be reiterated that the beauty of the whole procedure is
that a unique Lagrangian is able to describe in a unified way
$all$ hadronic interactions, the ground state and fluctuations
based on it. This hadronic theory provides us with results which
are promising and exciting while exposing weaknesses of the more
traditional approaches. Needless to say, there is still a long way
to go before confronting Nature, such as, for instance, the
properties of the pionic atoms that are offering a tantalizing
view of what is going on in dense medium and of course the
interior of compact stellar systems where chiral symmetry is
presumably restored. Properly and realistically formulated, we
might be able to ``derive" from the first principles such novel
notions as BR scaling~\cite{BR91,BR:PR01}, the hidden local
symmetry/vector manifestation~\cite{HY:PR} etc.

\section*{Acknowledgments}
Hee-Jung Lee, Byung-Yoon Park and Vicente Vento are grateful for
the hospitality extended to them by KIAS, where this work was
finished. This work was partially supported by grants
MCyT-BFM2001-0262 and GV01-26~(VV), KOSEF Grant R01-1999-000-00017-0~(DPM,
BYP), M02-2002-000-10008-0~(HJL), Brain Korea 21 Project in 2002,
and KRF Grant 2001-015-DP0085~(DPM).

\end{document}